\documentclass[preprint, onecolumn, showpacs, aps,prb,amssymb,amsmath]{revtex4}

\usepackage{amssymb}
\usepackage{amsmath}
\usepackage{graphicx}
\usepackage{subfigure}

\def\({\left(}
\def\){\right)}
\def\[{\left[}
\def\]{\right]}

\newcommand{\e}{\begin{equation}}
\newcommand{\q}{\end{equation}}
\newcommand{\m}{\begin{eqnarray}}
\newcommand{\n}{\end{eqnarray}}
\newcommand{\be}{\begin{equation}}
\newcommand{\ee}{\end{equation}}
\newcommand{\bea}{\begin{eqnarray}}
\newcommand{\eea}{\end{eqnarray}}

\begin{document}

\title{Resonant Tunneling in Superfluid Helium-3}

\author{S.-H. Henry Tye\footnote{sht5@cornell.edu, iastye@ust.hk} and Daniel Wohns\footnote{dfw9@cornell.edu}}
\affiliation{Laboratory for Elementary-Particle Physics, Cornell University, \\
 \hspace{0.5ex} Ithaca, NY 14853, USA \\
Institute for Advanced Study, The Hong Kong University of Science and Technology \\
 \hspace{0.5ex} Clear Water Bay, Hong Kong}

\date{\today}

\begin{abstract}
The $A$ phase and the $B$ phase of superfluid He-3 are well studied, both theoretically and experimentally. The decay time scale of the $A$ phase to the $B$ phase of a typical supercooled superfluid $^3$He-A sample is calculated to be $10^{20,000}$ years or longer, yet the actual first-order phase transition of supercooled $A$ phase happens very rapidly (in seconds to minutes) in the laboratory. We propose that this very fast phase transition puzzle can be explained by the resonant tunneling effect in field theory, which generically happens since the degeneracies of both the $A$ and the $B$ phases are lifted by many small interaction effects. This explanation predicts the existence of peaks in the $A \to B$ transition rate for certain values of the temperature, pressure, and magnetic field. 
Away from these peaks, the transition simply will not happen.

\end{abstract}


\maketitle

\section{Introduction}

It is well known that superfluid He-3 has a very rich phase structure \cite{Review}. Its superfluidity properties allow a typical sample to be treated as a pure quantum system devoid of impurities. A number of its phases have been well studied, in particular the $A$ phase and the $B$ phase. They are well described by the so-called mean field theory.  
Their properties (such as free energy density difference, critical temperature and domain wall tension \cite{Osheroff-Cross,Kaul, schopohl}) are well understood and measured so the $A \to B$ phase transition rate can be reliably calculated. At low enough temperature, $B$ phase has a lower free energy density than that of the $A$ phase. In the nucleation theory for a supercooled $A$ phase sample, the $A \to B$ first order phase transition can go via thermal fluctuations or via quantum tunneling. The characteristic time for a typical sample in $A$ phase (the false ground state) to thermally fluctuate over the barrier is \cite{Schiffer:1995zza,LY}
\e
\label{est1}
T \sim 10^{1,470,000} \, {\rm s}
\q
(Note that choosing the units in years instead of seconds
leads to a tiny error in the exponent, well within the uncertainties of the estimate.) If it goes via quantum tunneling at zero temperature, one obtains, in the usual WKB approximation \cite{BL},
\e
\label{est2}
T \sim 10^{20,000} \, {\rm s}
\q
This estimate at zero temperature is too optimistic for the actual situation.  At higher temperatures where the transition has been observed, the quantum tunneling time is estimated to be longer (the exponent is bigger by at least an order of magnitude).  These estimates imply that the transition should never have happened. Yet, it is a well known fact that this transition actually happens very rapidly, in hours if not in seconds. This very rapid transition allows experimentalists to reach and study the $B$ phase by supercooling the superfluid He-3 in the $A$ phase (which is in turn usually reached via the $A_1$ phase). The discrepancy between theory and experiments is huge : the above exponents are too big by four to five orders of magnitude. 

Superfluid He-3 is one of the most pure quantum systems accessible in the laboratory.  Any impurity will self separate (e.g., He-4 will sink to the bottom). So superfluid He-3 is an excellent quantum system to study. It is intuitively clear that impurities can provide seeds of nucleation bubbles for the transition. Since superfluid He-3 has no impurities, external beams such as cosmic rays may provide the necessary seeds of nucleation, thus exponentially speed up the phase transition process. This is the ``Baked Alaska" model \cite{baked}.  Alternatively the ``cosmological" scenario \cite{BunkovTim} proposes that after a local heating above the superfluid transition temperature, many casually independent regions undergo the superfluid transition to the $A$ or $B$ phase.  If the energetically favorable $B$ phase seeds percolate, the transition will complete.  Although we agree that external interference (e.g., shooting neutron beams or cosmic rays on the sample) can surely speed up the transition process \cite{Schiffer:1995zza}, a direct search of cosmic ray effect detected no such correlation in a superfluid He-3 sample \cite{Swift,LANL}.  So this observed superfast phase transition remains an outstanding puzzle. Here, we propose to explain this rapid phase transition as a natural consequence of the resonant tunneling phenomenon.

If our explanation is correct, there is at least a plausible, qualitative but very distinctive prediction that may be readily checked experimentally. Resonant tunneling phenomenon happens only under some fine-tuned conditions. This feature predicts the existence of peaks in the $A \to B$ transition rate for certain values of the temperature, pressure, and external magnetic field.
Away from these peaks, the transition simply will not happen. These high probability regions may take the form of isolated peaks, or lines or surfaces in the three-dimensional space with temperature, pressure, and magnetic field as the three coordinates. The locations and shapes of such regions should also depend on the container geometry as well as the properties of the container surface.

Experiments in Ref.\cite{LANL,Hakonen} have shown that, for fixed pressure, magnetic field and geometry, the $B$ phase nucleation takes place at a specific temperature. For example, Ref.\cite{Hakonen} finds that, for pressure at $29.3$ bar and magnetic field $H=28.4$ mT, $B$ phase nucleation takes place at temperature $T =0.67T_c$ with a full width of about $0.02T_c$ (where $T_c$ is the superfluid transition temperature) when the sample is slowly cooled, i.e., there is a peak in the plot of transition event number vs $T$. This is what our proposal expects : the resonant tunneling condition is satisfied only when the properties of the He-3 sample are just right. This happens at a specific temperature when other conditions are fixed. Now, the resonant condition is simply the Bohr-Sommerfeld quantization condition (\ref{resonancecond}), which has multiple solutions. This allows the possibility that there are more than one nucleation temperature. In the three-dimensional space with pressure, magnetic field and temperature as the three coordinates, there are isolated regions where the $A \to B$ transition is fast enough to be observed.  This also suggests the following two possibilities \cite{Hakonen} : \\
(1) The resonant peak in the event number (of $B$ phase transition) versus temperature is actually an unresolved collection of two or more extremely narrow peaks.  \\
(2) The width may be due to the spread caused by the finite temperature and experimental setup resolution limit.
Depending on the details, there may be other critical nucleation temperatures besides the one observed. A simple search of additional nucleation temperatures below $T =0.67T_c$ will be very interesting. If they exist, we expect their widths to be narrower as well.

In quantum mechanics (QM) the tunneling probability (or the transmission coefficient) of a particle incident on a barrier is typically exponentially suppressed.  However, the addition of a second barrier can actually enhance the tunneling probability: under appropriate conditions and for specific values of the particle's energy, the tunneling probability may actually approach unity.  This enhancement in the tunneling probability, known as resonant tunneling, is due to the constructive interference of a set of quantum paths of the particle through the barriers. This is a very well understood phenomenon in quantum mechanics \cite{Merzbacher,Tye:2006tg}. The first experimental verification of this phenomenon was the observation of negative differential resistance due to resonant tunneling in semiconductor heterostructures \cite{CET}. In fact, this phenomenon has industrial applications, e.g., resonant tunneling diodes etc. \cite{wiki}.

Tunneling under a single barrier in quantum field theory (QFT) with a single scalar field is well understood, following the work of Langer, Coleman and others on the formation of nucleation bubbles \cite{Langer,Coleman:1977py}.
Using the functional Schr\"odinger method \cite{Gervais:1977nv}, one can show how the resonant tunneling phenomenon through double barriers takes place in quantum field theory with a single scalar field \cite{TyeWohns}. Again, this phenomenon can lead to an exponential enhancement of the single-barrier tunneling rate. To identify this resonant tunneling phenomenon in nature, we need a quantum system with multiple false vacua with appropriate properties. In this paper, we present arguments that this resonant tunneling phenomenon has already been observed in superfluid He-3.

The properties of the $A$ and $B$ phases are among the best understood, both theoretical and experimental, qualitative and quantitative, in condensed matter physics. The He-3 pairing is in $p$-wave spin $S=1$ state. Here the order parameter (a $3 \times 3$ complex matrix ${\bf \Delta}({\bf r},t)$) describes the properties of various phases and the real scalar field $\phi$ we have in mind is the interpolating field among the specific phases the particular transition is taking place.  Although the calculated thermal fluctuation time (\ref{est1}) is faster than single-barrier quantum tunneling time at the temperatures at which the transition is observed, under appropriate conditions the resonant tunneling effect can reduce the exponent in Eq.(\ref{est2}) by orders of magnitude, reversing the inequality. (Pre-factors will be ignored throughout.)

In this paper, we present the relevant conditions for the resonant tunneling phenomenon and propose how it may happen in the $A \to B$ phase transition. It is well accepted that both the $A$ phase and the $B$ phase actually consist of multiple distinct local classically metastable minima of the free energy functional, which we shall refer to as $A$ sub-phases and $B$ sub-phases. To avoid confusion with the $A_1$ phase, we shall refer to these as $A^i$ sub-phases (similarly for the $B$ sub-phases). The barrier and the free energy density difference between any two $A$ sub-phases are small compared to that between an $A$ sub-phase and a $B$ sub-phase (similarly for the $B$ sub-phases). Here, we start with this qualitative property and show that the $A^i \to A^j \to B$ transition can easily be enhanced by the resonant tunneling effect. This enhancement can be particularly strong for a specific $B^k$ sub-phase, so the $A^i \to A^j \to B^k$ transition will dominate. Even though we may not know the actual sub-phases involved in a specific sample, we argue that this phenomenon is quite generic in He-3. As supercooling is taking place, the detailed properties of the sub-phases are slowly changing accordingly, increasing the probability of hitting  the resonance condition at certain point, and so a typical transition can be quite fast.

Our analysis further suggests that the resonant tunneling effect may remain as both the
$A^i-A^j$ tension $\sigma_{1}$ and the $A^i-A^j$ energy density difference $\epsilon_{1}$ approach zero while keeping the ratio $\sigma_{1}^4/(\epsilon_{1}^3 v_F \hbar)$ large but fixed (where $v_F$ is the Fermi velocity). This suggests that the resonant effect may remain when we have a degenerate or an almost degenerate $A$ phase. Here the resonant tunneling effect probably follows from the coherence of the infinite sum of Feynman paths in the degenerate $A=A^i-A^j$ phase. Further study will be important in finding the necessary condition for resonant tunneling in this case.

We shall also compare this resonant tunneling scenario to other proposed explanations to this fast transition puzzle and discuss some possible ways to test this proposal.

Our original motivation to study resonant tunneling is its possible implication in string theory and cosmology \cite{Tye:2006tg,Tye:2007ja}. String theory suggests a multi-dimensional ``landscape" with numerous (if not infinite number of) classically stable local vacua (i.e., phases) \cite{KKLT}.
Tunneling between possible vacua in this cosmic landscape is an outstanding problem under investigation.
A better understanding of the first order phase transition processes in superfluid He-3 will certainly help, since the actual tunneling processes are rather complicated, so it is truly useful that one can do experiments to test the model calculations. In this sense, this is another way to realize the connection of He-3 to cosmology \cite{Volovik}. It will be very useful to find other systems in the laboratory that exhibit the resonant tunneling phenomenon.

The rest of the paper contains the following sections. In Sec. 2, we review some of the properties of phase transitions in superfluid He-3 that are relevant to the above estimates of the transition rates (\ref{est1},\ref{est2}). This brings out clearly the puzzle. In Sec. 3, we review the formalism for resonant tunneling in scalar field theory \cite{TyeWohns}. In Sec. 4, we study the conditions for resonant tunneling and identify the tunneling $A^i \to A^j \to B^k$ to be most likely, as compared to other transitions that involve an intermediate $B$ sub-phase. ${\it A}$ ${\it priori}$, other tunneling paths, say those involving other possible phases in superfluid He-3, may be potential candidates too. In Sec. 5 we discuss some possible ways to test this proposal.  Sec. 6 contains some remarks. The appendix reviews the resonant tunneling effect in quantum mechanics.

\section{Thermal and Quantum Tunneling} \label{thermalquantum}

The features of superfluid He-3 physics is well described by the mean field theory. Both the $A$-phase and the $B$-phase are continuously degenerate. These degeneracies are typically lifted by the presence of external magnetic field, which interacts with the spin and the orbital waves, the container wall effect, as well as the wall surface irregularities etc.. Let us ignore these effects for the moment and consider the tunneling between the $A$-phase and the $B$-phase. Here we like to review the inputs that go into the estimates (\ref{est1}, \ref{est2}) and show that these estimates of the exponents are reasonable within the present context.

A He-3 atom has spin one-half and the He-3 pairing happens in the spin $S=1$ $p$-wave state. So the order parameter is a $3 \times 3$ matrix $\Delta_{\alpha i}$, where $\alpha \in (x,y,z)$ is the spin index and $i \in (x,y,z)$ is the index
for the ${\it l}=1$ $p$-wave orbital.

Assuming that the order parameter takes the shortest path in field space from the $A$-phase to the $B$-phase, the order parameter takes form
\begin{equation}
\label{path}
\Delta_{\alpha i} = \frac{\Delta(A)}{\sqrt{2}}(1-\zeta)  \left( \begin{array}{ccc}
1 & i & 0 \\
0 & 0 & 0 \\
0 & 0 & 0 \end{array} \right) +  \frac{\Delta(B)}{\sqrt{3}} \zeta \left( \begin{array}{ccc}
1 & 0 & 0 \\
0 & 1 & 0 \\
0 & 0 & 1 \end{array} \right)
\end{equation}
for the configuration of interest. The false ground state, namely the $A$ phase, is at $\zeta=0$ and the true ground state, namely the $B$ phase, is at $\zeta=1$. Up to a normalization factor, $\zeta$ is simply the interpolating field $\phi$.

Although we shall not go into any details, it is important to point out the following key point. It is obvious from the form of the $A$ phase matrix that it is highly degenerate. For example, instead of putting the non-zero values in the $xx$ and $xy$ entries, we can rotate them into other entries. Besides the standard Ginzburg-Landau free energy functional, there are many other interaction terms that will contribute to the free energy density \cite{Review}. Some examples include interactions with the external magnetic field and the container wall.  Magnetic field effects are generally small, but container wall (which typically can have some irregularities on its surface) effects can be very strong for He-3 close to the wall. There are texture and topological properties, as well as current properties. In general, these effects tend to lift (or reduce) the large degeneracy of the $A$ phase, leading to many $A$ sub-phases. A similar situation happens for the $B$ phase. This fact will play a crucial role in our proposal.

The critical temperature depends on the pressure and the magnetic field. Here we shall use the typical value  $T_c \approx 2.5 \rm{mK}$. The measured value of the domain wall tension between the $A$ and $B$ phases at melting pressure is $\sigma \simeq 9.3 \times 10^{-9} \rm{J}/\rm{m}^2$ \cite{Osheroff-Cross} while the calculated value using the path (\ref{path}) in the Ginzburg-Landau free energy functional is within 10\% of the measured value \cite{Kaul}.  For a more accurate calculation of the domain wall tension see \cite{schopohl}.  At $T=0.7T_c$ (a typical temperature in the experiments), the free energy density difference is $\epsilon = 0.013 \rm{J}/\rm{m}^3$ \cite{Review,Schiffer:1995zza}.
Note that both $\sigma  \propto  \left( 1-{T}/{T_c} \right)^{1/2} $ and $\epsilon \propto  \left( 1 - {T}/{T_c} \right)^2$ are temperature dependent.

The decay width of the $A$ phase to the $B$ phase is given by, ignoring the prefactor,
\e
\Gamma \simeq e^{-S / \hbar}
\end{equation}
Using the above values for $\sigma$ and $\epsilon$ (at $T=0.7T_c$), the exponent  $S$ for the pure quantum tunneling decay process in the thin-wall approximation is given by \cite{Langer,Coleman:1977py}
\begin{equation}
\label{Squantumsimple}
S_{\rm{quantum}} = \frac{27 \pi^2}{2} \frac{\sigma^4}{\epsilon^3}\frac{1}{v_F}= 8.2 \times 10^7 \hbar
\end{equation}
 if the Fermi velocity $v_F$ is about $55 \rm{m}/{\rm s}$. 
 The difference between this estimate and the estimate (\ref{est2}) is in the values of $\sigma$ and $\epsilon$ used. The estimate (\ref{est2}) uses instead the values of $\sigma$ and $\epsilon$ at $T=0$ while the actual  temperatures in the experiments are closer to the value ($T=0.7T_c$) we use.

As a simple estimate of the validity of the thin-wall approximation, we can compute the ratio of the radius of the bubble at nucleation $\lambda_c$ to the thickness of the domain wall $1/\mu$.  For a symmetric double well potential (ignoring the small $\epsilon$ term), $\mu \approx \sqrt{2} m v_F / \hbar$, where $m$ is the mass of the scalar field in the false vacuum.  The correlation length $\xi_0$ is of order $\hbar/m v_F$, and $\xi_0 = (\frac{7\zeta(3)}{48 \pi^2})^{1/2} \frac{\hbar v_F}{k T_c}$ implies $\xi_0 \approx 15 {\rm nm}$, so $1/\mu \sim 10 {\rm nm}$.  In the thin-wall limit, $\lambda_c = 3 \sigma / \epsilon \approx 2000 {\rm nm}$, so the radius of the bubble is $\mathcal{O}(100)$ times the thickness of the domain wall, and the approximation is consistent.

The simplest estimate one could do for the pure thermal activation exponent simply uses the Boltzmann factor:
\begin{equation}
\label{Sthermalsimple}
S_{\rm{thermal}} = \frac{16 \pi}{3} \frac{\sigma^3}{\epsilon^2} \frac{1}{k_B T}=3.3 \times 10^6 \hbar
\end{equation}

The actual tunneling takes place via a combination of quantum and thermal processes.
In a scalar quantum field theory with a false vacuum and a true vacuum, tunneling starts from the bottom of the false vacuum in the potential.
At finite temperature, tunneling does not need to occur from the bottom
of the false potential well.  Instead, tunneling proceeds by a
combination of thermal excitation part way up the
barrier followed by quantum tunneling through the barrier.



More detailed estimates \cite{Schiffer:1995zza,BL} using different interpolations agree with these simple estimates of $S$ to within a factor of $\mathcal{O}(10)$. In any case, although the actual estimate of $S$ may vary somewhat, no reasonable theoretical argument can push the value of $S$ substantially below that of Eq.(\ref{est2}), which implies that the $A \to B$ phase transition should never have happened. This is the puzzle we are facing.




\section{Review of Resonant Tunneling} \label{Review}

In \cite{TyeWohns}, the resonant tunneling effect for a scalar quantum field theory with a triple-well potential in the double thin-wall limit was established.
Let us start with a scalar quantum field theory with the interpolating field $\phi ({\bf x})$ and its effective potential $V(\phi ({\bf x}))$. The time-independent functional Schr\"odinger equation is \cite{Gervais:1977nv}
\begin{equation}
H \Psi(\phi({\bf x})) = E \Psi(\phi({\bf x}))
\end{equation}
where
\e
H = \int d^3{\bf x} \bigg(- \frac{{\hbar}^2}{2}\bigg( \frac{\delta}{\delta \phi ({\bf x})}\bigg)^2+ \frac{1}{2} (\nabla \phi)^2 +V(\phi)\bigg) \,\,,
\q
and the eigenvalue $E$ is the energy of the system. As usual $\Psi(\phi({\bf x}))= A \exp (-\frac{i}{\hbar}S(\phi))$ is the amplitude that gives a measure of the likelihood of the occurrence of the field configuration $\phi({\bf x})$.


At leading order in $\hbar$ the functional Schr\"odinger equation reduces to a one-dimensional WKB equation.  The key idea that allows this simplification is that there is a trajectory in the configuration space of $\phi({\bf x})$ known as the most probable escape path (MPEP) which provides the dominant contribution to the tunneling probability \cite{Banks:1973ps}.  The MPEP $\phi_0({\bf x}, \lambda)$ satisfies the Euclidean equations of motion in the classically forbidden regions $U(\phi_0({\bf x}, \lambda))>E$, and satisfies the Lorentzian equations of motion in the classically allowed regions $U(\phi_0({\bf x}, \lambda)) < E$, where the effective tunneling potential $U(\phi_0({\bf x}, \lambda))$ is
\begin{equation}
\label{Ul}
U(\lambda) = U(\phi_0({\bf x}, \lambda)) = \int {\rm d}^3{\bf x} \left( \frac{1}{2} (\nabla \phi_0({\bf x}, \lambda))^2 + V(\phi_0({\bf x}, \lambda))\right).
\end{equation}
In general $\lambda$ is an arbitrary real parameter.
Since we are integrating over the three spatial dimensions, it is convenient to choose $E=0$ when sitting at the bottom of the highest false vacuum so  $U(\lambda)$ will stay finite.
At leading order in $\hbar$ the functional Schr\"odinger equation reduces to a one-dimensional time-independent Schr\"odinger equation:
 \e
 \label{QMl}
\bigg( - \frac{\hbar^2}{2} \frac{d^2}{d\lambda^2} +m(\lambda)U(\lambda) \bigg) \Psi_0 (\lambda) =0
 \q
 where $m(\lambda)$ is the effective mass, which is manifestly positive,
 \e
\label{m}
m(\lambda) \equiv \int d^3x \bigg( \frac{\partial \phi_0 ({\bf x}, \lambda)}{\partial \lambda}\bigg) ^2
\q
Here $\lambda$ plays the role of a spatial coordinate. Given $V(\lambda)=m(\lambda)U(\lambda)$, Eq.(\ref{QMl}) can be readily solved. For the single barrier case, as expected, the WKB result of Eq.(\ref{QMl}) reproduces Coleman's instanton result in the thin-wall approximation.

 \begin{figure}
\begin{center}
\includegraphics[width=0.4\textwidth]{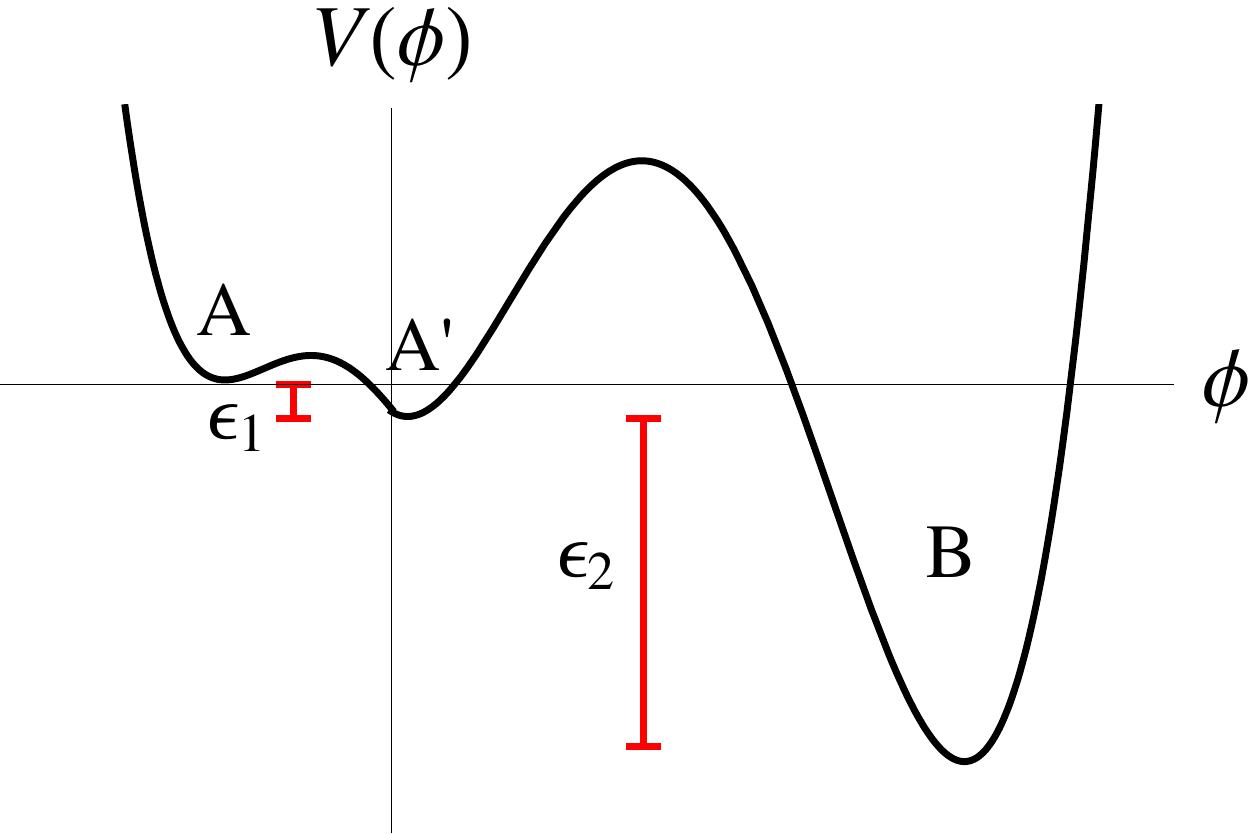}
\caption{A typical effective potential $V(\phi)$ with a false vacuum $A$ at $\phi=-2c_1$, a slightly lower (by $\epsilon_1$) false vacuum $A'$ at $\phi=0$ and  a true vacuum $B$ at $\phi=2c_2$.}
\label{double_barrierV}
\end{center}
\end{figure}

Let us consider the double-barrier case. Starting with a potential $V(\phi)$ as shown in Fig. \ref{double_barrierV}, we use the intuitively obvious MPEP for $\phi({\bf x},t)$ to tunnel from the $A$ phase to the $B$ phase via the $A'$ phase : sitting at the false vacuum $A$, a nucleation bubble with $A'$ phase inside starts to form. In the region in the $A'$ phase, another bubble with $B$ phase inside starts to form.
To simplify the problem, we shall consider the (double) thin-wall approximation. As a function of the four-dimensional radial coordinate $r$, we have the MPEP
\e
\label{gensol}
\phi(r) = -c_1 \tanh \bigg( \frac{\mu_1}{2} (r-r_1) \bigg) - c_2 \tanh \bigg( \frac{\mu_2}{2} (r-r_2) \bigg) +c_2-c_1
\q
where $1/\mu_{1,\,2}$ measure the thicknesses of the domain walls.
For appropriate $r_1 > r_2$, we shall use this $\phi$ (\ref{gensol}) as an ansatz to find the resonant tunneling condition.

It is straightforward to extract $\phi({|x|}, \lambda)$ from $\phi (r)$ given by Eq.(\ref{gensol}),
\begin{eqnarray}\label{MPEPRT}
\phi_0({|x|},\lambda) = -c_1 \tanh \bigg( \frac{\mu_1}{2}\frac{\lambda}{r_1} ({|x|}-\lambda) \bigg)
-\nonumber \\ \Theta\bigg(\frac{\lambda }{\Lambda} - 1\bigg) c_2 \tanh \bigg( \frac{\mu_2}{2}\frac{\lambda'}{r_2} ({|x|} - \lambda') \bigg) +c_2-c_1
\end{eqnarray}
where we use 
\begin{equation}
\lambda = \left\{
    \begin{array}{lcl}
      \sqrt{\lambda_c^2-\tau^2}  & :&  \textrm{ classically forbidden}\\
    \sqrt{\lambda_c^2+t^2} &:&     \textrm{ classically allowed}
     \end{array}
   \right.
\end{equation}
Here $\Lambda$ is the value of $\lambda$ at which the inside bubble has zero spatial extent,
\e
\label{geometry_constraint}
\Lambda ^2 = r_1 ^2 - r_2^2
\q
and as long as both bubbles are expanding
\e
\label{lambdaprime}
\lambda ^\prime = \sqrt{\lambda ^2 - \Lambda^2} \,\,.
\q

 \begin{figure}
\begin{center}
\includegraphics[width=0.25\textwidth]{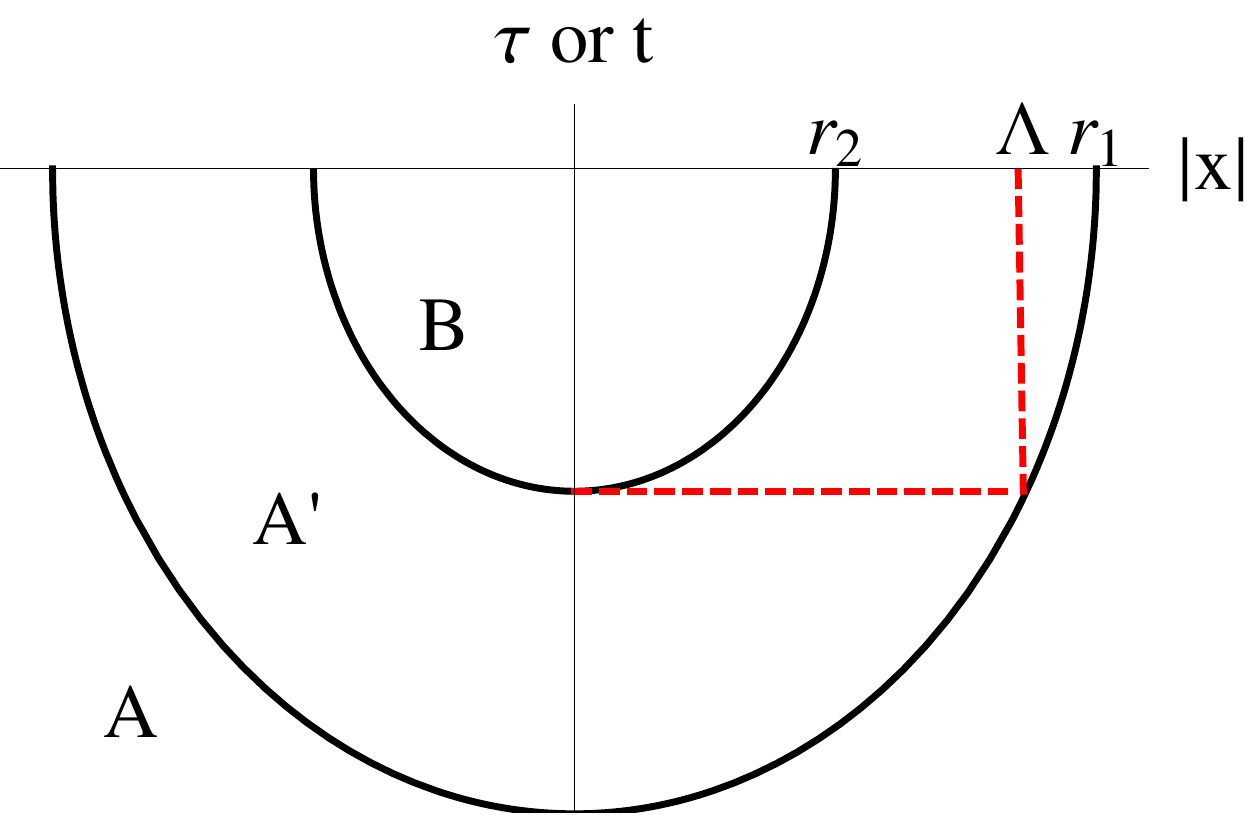}
\caption{The tunneling process from $A$ to $B$ via $A'$ leads to the formation of two bubbles: the outside bubble separates $A'$ from $A$ and the inside bubble separates $B$ from
$A'$. They are drawn as concentric bubbles here, though the inside bubble can shift sideways.
Here $\Lambda$ is the length of the horizontal dashed (red) line. We have $r_1 > \Lambda > r_2$.}
\label{bubbles}
\end{center}
\end{figure}

This is shown in Figure \ref{bubbles}. The equation (\ref{gensol}) also implies that $\phi_0({|x|}, \lambda)=0$ for $\lambda <0$.
Substituting this MPEP $\phi_0({|x|}, \lambda)$ given by Eq.(\ref{MPEPRT}) into Eq.(\ref{Ul}) and Eq.(\ref{m}) now yields, after a straightforward calculation, the effective mass $m(\lambda)$ and the effective tunneling potential
$U(\lambda)= U(\phi_0({|x|},\lambda))$.
 In the double thin-wall limit, the potential is described by the domain wall tensions $\sigma_1$ (between $A$ and $A'$) and $\sigma_2$ (between $A'$ and $B$) and the energy density differences $\epsilon_1$ and $\epsilon_2$.  In this limit the effective tunneling potential is
 \begin{eqnarray}
\label{U_double}
U(\lambda) =
2 \pi \sigma_1 \bigg( \frac{\lambda}{r_1} + \frac{r_1}{\lambda} \bigg) \lambda^2 - \frac{4 \pi}{3} \epsilon_1 \lambda^3
+\nonumber \\
 2 \pi \sigma_2 \bigg( \frac{\lambda^ \prime }{r_2} + \frac{r_2}{ \lambda ^\prime} \bigg) (\lambda^\prime)^2 -
\frac{4 \pi}{3} \epsilon_2 (\lambda^ \prime)^3
\end{eqnarray}
where $r_1$ and $r_2$ are the radii of the larger and smaller bubbles respectively upon nucleation, $\Lambda^2 = r_1^2 - r_2^2$, and
\begin{equation}
\lambda^\prime = \left\{
     \begin{array}{lcl}
       \sqrt{\lambda^2-\Lambda^2} &:& \Lambda < \lambda\\
       0 &:&   \textrm{otherwise}
     \end{array}
   \right.
\end{equation}
is the spatial radius of the smaller bubble.  The position-dependent mass is
\begin{equation}
m(\lambda) = 4 \pi \bigg( \frac{\sigma_1}{r_1}\lambda^2 + \Theta(\lambda/\Lambda - 1)\frac{\sigma_2 \lambda}{r_2 \lambda^
\prime}(\lambda^\prime)^2\bigg) \lambda \,\,.
\end{equation}

Conservation of energy implies that the radii of the two bubbles when they simultaneously nucleate and begin to grow classically are related via
\begin{equation}
\label{energy}
E = U(r_1) = 4 \pi (\sigma_1 - \frac{1}{3} r_1 \epsilon_1) r_1^2 + 4 \pi (\sigma_2 - \frac{1}{3}r_2 \epsilon_2) r_2^2 = 0 \,\,.
\end{equation}
We restrict our attention to zero-energy tunneling which implies the last equality in (\ref{energy}).  After imposing the constraint (\ref{energy}), we still have one free parameter, which we can take to be $r_2$.

\begin{figure}
\centering
\mbox{\subfigure{\includegraphics[width=1.4in]{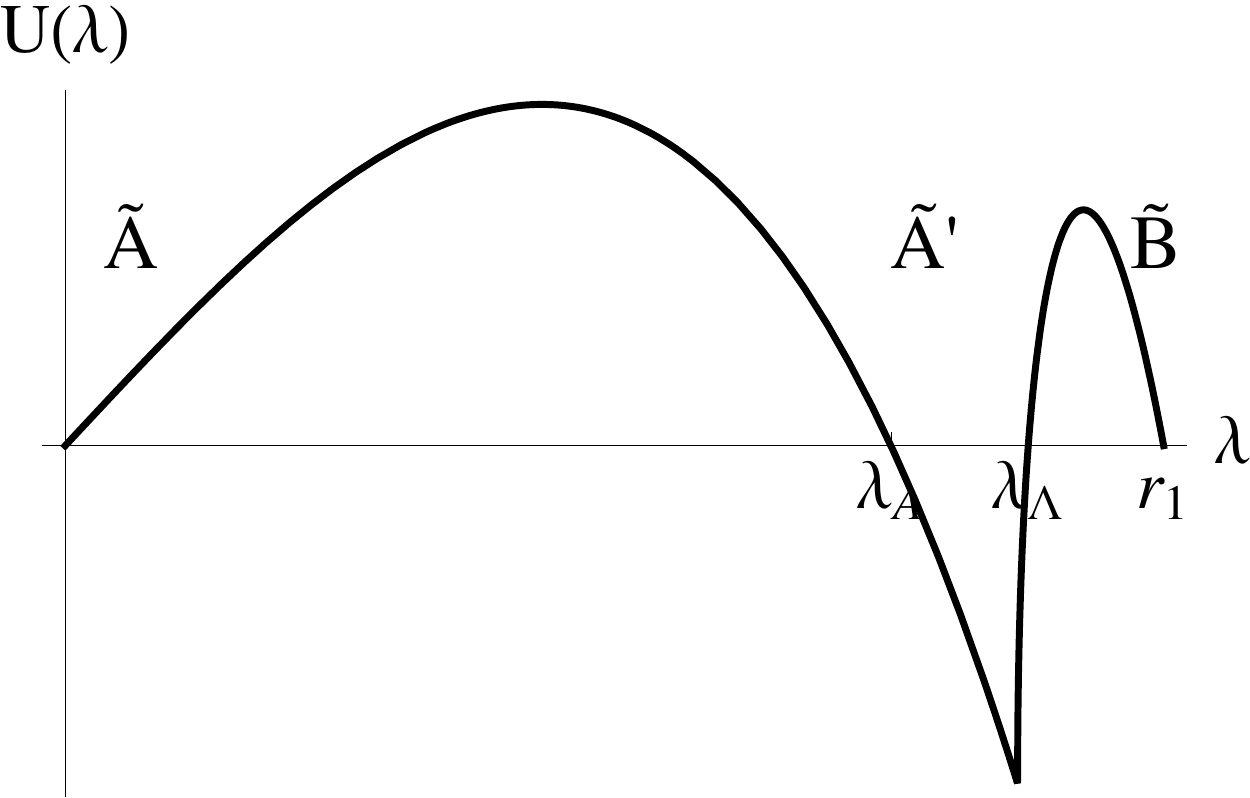}
\quad
\subfigure{\includegraphics[width=1.4in]{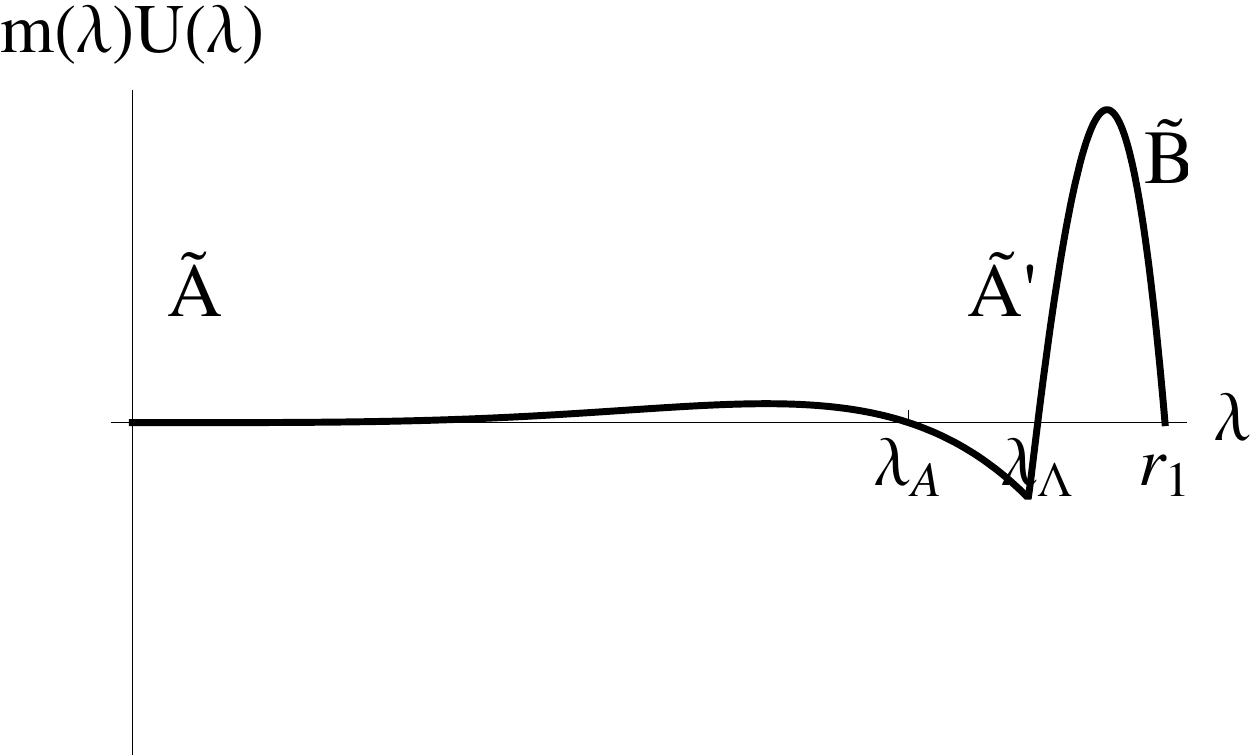} }}}
\caption{Left : $U(\lambda)$ for the double barrier potential $V(\phi)$ in Fig. \protect \ref{double_barrierV}.
Here, $U(\lambda)=0$ for $\lambda <0$.
Right : $V(\lambda)=m(\lambda)U(\lambda)$ for the same double barrier case.}
\label{VmU}
\end{figure}

\begin{figure}
\begin{center}
\includegraphics[width=0.4\textwidth]{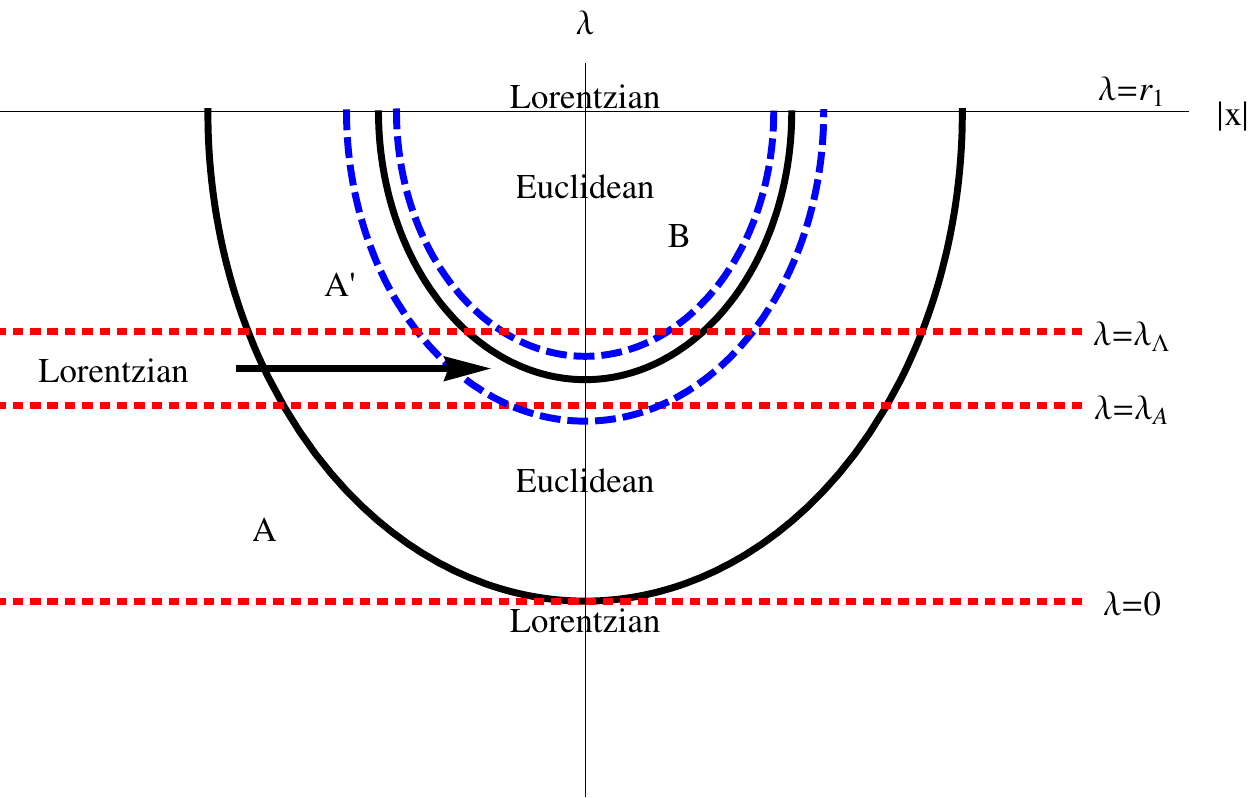}
\caption{The various regions that can be described by Euclidean time or by Lorentzian time in the double bubble nucleation process. This figure corresponds to a case that permits resonant tunneling. The boundaries between these regions are the line $\lambda = r_1$, the outer bubble wall, and the region around the inner bubble wall enclosed by the two blue dashed lines. The actual (semi-circular) region of Lorentzian time description shrinks slightly as we approach $\lambda=r_1$. In the leading order approximation in the functional Schr\"odinger method, the boundaries between these regions
are given by the red horizontal dotted lines and the line $\lambda = r_1$.}
\label{EuMin}
\end{center}
\end{figure}

However $r_2$ is subject to additional constraints, due to the necessity of having an effective tunneling potential $U(\lambda)$ capable of supporting resonant tunneling.  In particular the effective tunneling potential must have two distinct classically forbidden regions separated by a classically allowed region, and must energetically favor the growth of at least one bubble after nucleation.  Equivalently $U(\lambda)$ must have four zeros $0$, $\lambda_A$, $\lambda_\Lambda$, and $r_1$, and must approach negative infinity for large $\lambda$,
\e
\label{energy_constraint}
0 = U(0)= U(\lambda_{A}) = U(\lambda_{\Lambda}) = U(r_1) \,\,.
\q
The radii of the two bubbles at the moment of nucleation are related via the constraint (\ref{geometry_constraint}). This is illustrated in Figure \ref{EuMin}.
The determination of the various approximate Euclidean/Lorentzian regions is possible only after we determine the MPEP. {\it A} {\it priori}, it is difficult to determine the existence of the classically allowed region and evaluate the sum of the set of coherent Feynman paths before the problem is reduced to a ``time"-independent one-dimensional QM problem. This is why the functional Schr\"odinger method is very useful here, since it
completely avoids the introduction of either Euclidean time or real time into the tunneling framework.

The requirement that there are four distinct classical turning points leads to the conditions
\begin{eqnarray}
\label{twobarriers}
\Lambda^2 > \lambda_A^2 = \frac{\lambda_{1c}r_1^2}{2r_1 - \lambda_{1c}} \\ \nonumber
2 (\sigma_1 + \sigma_2) < r_1 \epsilon_1 + r_2 \epsilon_2
\end{eqnarray}
where $\lambda_{1c,\,2c}$ are the radii at which bubbles of $A^\prime$ or $B$ are nucleated in single-barrier tunneling processes.
The condition that at least one bubble must grow classically after nucleation is
\begin{eqnarray}
\label{expansion}
\frac{\sigma_1}{r_1} + \frac{\sigma_2}{r_2} < \frac{2 \epsilon_1}{3} + \frac{2 \epsilon_2}{3} \,\,.
\end{eqnarray}
For some choices of $\sigma_1$, $\sigma_2$, $\epsilon_1$, and $\epsilon_2$, these conditions are incompatible and preclude any possibility of a resonance.  In particular Eq.(\ref{twobarriers}) and Eq.(\ref{expansion}) rule out any possibility of a resonance effect for $\epsilon_1 < 0$ or $\epsilon_2 < 0$.

\begin{figure}
\begin{center}
\includegraphics[width=0.45\textwidth]{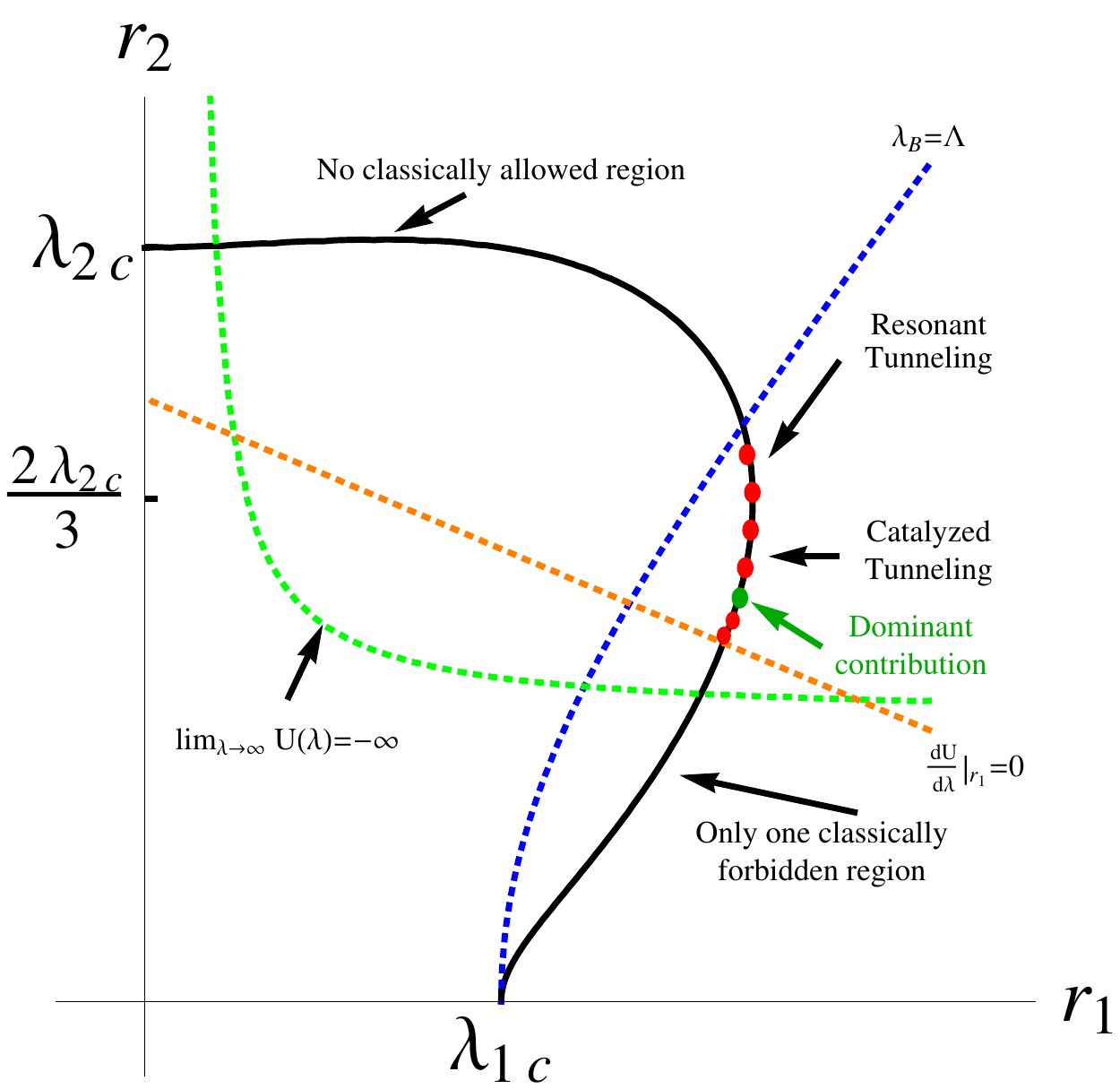}
\caption{The allowed radii $r_1$, $r_2$ of the bubbles are constrained to lie on the solid black curve by energy conservation.  The dashed lines show the boundaries of the constraints (\protect\ref{twobarriers}) and (\protect\ref{expansion}) that the effective tunneling potential (\protect\ref{U_double}) has four distinct classical turning points and at least one bubble can grow after nucleation.  For some choices of $\sigma_{1,2}$ and $\epsilon_{1,2}$, a finite number of points satisfy the resonant condition (\protect\ref{resonancecond}).  These points are indicated by dots.  If there is more than one such point, one will provide the dominant contribution to the tunneling probability.  Depending on the radius $r_2$ of the smaller bubble of this dominant contribution, either resonant tunneling from $A$ to $B$ or catalyzed tunneling from $A$ to $A^\prime$ will occur.}
\label{nconstraint}
\end{center}
\end{figure}

If $r_2 > 2 \sigma_1 / \epsilon_1$, the smaller bubble will grow classically after nucleation and the tunneling from $A$ to $B$ will complete.  We call this case resonant tunneling.  Alternatively if $r_2 < 2 \sigma_1 / \epsilon_1$, the smaller bubble will (classically) collapse after nucleation and the result will be tunneling from $A$ to $A^\prime$, although it is possible for $P(A \to A^\prime)$ to be much larger than $\exp (-S_E^{A \to A^\prime}/\hbar)$ where $S_E^{A \to A^\prime}$ is the Euclidean action of the $A$ to $A^\prime$ instanton.  We call this case catalyzed tunneling.

\section{Condition for Resonant Tunneling}

Now we are ready to find the condition for the resonant tunneling phenomenon. As is well known, the mean field theory allows many possible phases for superfluid He-3. So far only the normal, the $A_1$, the $A$ and the $B$ phases have been observed, where the last three are superfluids. The $A_1 \to A$ phase transition is second order, while the transition from $A$ to $B$ is first order. As we have mentioned in Section \ref{thermalquantum}, theoretical calculations argues that the $A$ to $B$ transition should never have happened, which is frequently contradicted by experiments. In a normal condensed matter system, impurities are typically present and they can provide seeds of bubble nucleation. However, since He-3 is devoid of impurities, the answer should lie somewhere else. It is possible that cosmic rays hitting the sample may play a role, as proposed in the ``Baked Alaska" model \cite{baked,LY} and the ``cosmological" scenario \cite{BunkovTim}. Showering beams of particles or ionizing radiation should certainly enhance the transition rate \cite{Schiffer:1995zza}. However, there is strong evidence that the transition rate is puzzlingly fast even in the absence of any such external disturbances. We believe the fast transition is a resonant tunneling phenomenon.

To apply the resonant tunneling phenomenon to the $A \to B$ transition, there are {\it a priori} two possibilities : via catalyzed tunneling or resonant tunneling. In both cases, a third phase besides $A$ and $B$ must be present in the phase diagram. At first sight, one may consider one of the predicted but as yet not discovered phases of He-3. However, none of them seems to have the right properties. Our analysis shows that resonant tunneling is the likely scenario and that requires a phase between the $A$ and the $B$ phases but very close to the $A$ phase; that is, both the domain wall tension $\sigma_1$ and the free energy density difference $\epsilon_1$ between that phase and the $A$ phase should be small compared to that ($\sigma_2$ and $\epsilon_2$) between that phase and the $B$ phase. How can we find such a phase?

As is well known, both the $A$ phase and the $B$ phase in a superfluid He-3 model are degenerate. These degeneracies are typically weakly lifted by the presence of an external magnetic field, and by the spin-orbit interaction. Furthermore, container wall effect has a large impact on the ground states of superfluid He-3 close to the wall. So both the $A$ phase and the $B$ phase are actually a collection of phases. Some of these phases have rich intricate properties; they are well studied, both theoretically and experimentally. Depending on the experimental setup and conditions, the initial $A$ phase is actually in one of these sub-phases. Let us call it the $A^i$ phase. If there is another $A$ sub-phase which has a lower free energy density than that of $A^i$, then we are in business. We call this sub-phase the $A^j$ phase.

With resonant tunneling, the decay rate of $A^i \to A^j \to B$ will be substantially enhanced. However, even with resonant tunneling, this decay rate is typically still exponentially suppressed (see Eq.(\ref{resonancetnn})). For example, even if the exponent in Eq.(\ref{est2}) is reduced by a factor of 100, the decay time is probably still far too long. Here, the presence of $B$ sub-phases should help. Again the various $B$ sub-phases have slightly different free energy densities and the domain wall tension $\sigma_2$ between $A^j$ and a $B$ sub-phase varies a little from one $B$ sub-phase to another. That is, nature will pick the particular $B$ sub-phase, called the $B^k$ sub-phase, that has the fastest tunneling rate. That is, the choice of a specific $B$ sub-phase provides a fine-tuning to enhance further the tunneling rate. Again, this enhancement is in the exponent for the $A^i \to A^j \to B^k$ transition.

As noted in section \ref{Review}, fixing the potential $V(\phi)$ and imposing energy conservation does not uniquely fix the size of the bubbles.  A range $r_{2,\rm{min}} < r_2 < r_{2,\rm{max}}$ of bubble sizes is determined via energy conservation (\ref{energy_constraint}) and the constraints (\ref{twobarriers}) and (\ref{expansion}).  Because $r_2$ is not fixed uniquely by these constraints, the tunneling probability $P(A \to B)$ (\ref{A_to_C}) depends nontrivially on the parameters $\sigma_{1,2}$ and $\epsilon_{1,2}$.

\begin{figure}
\begin{center}
\includegraphics[width=0.45\textwidth]{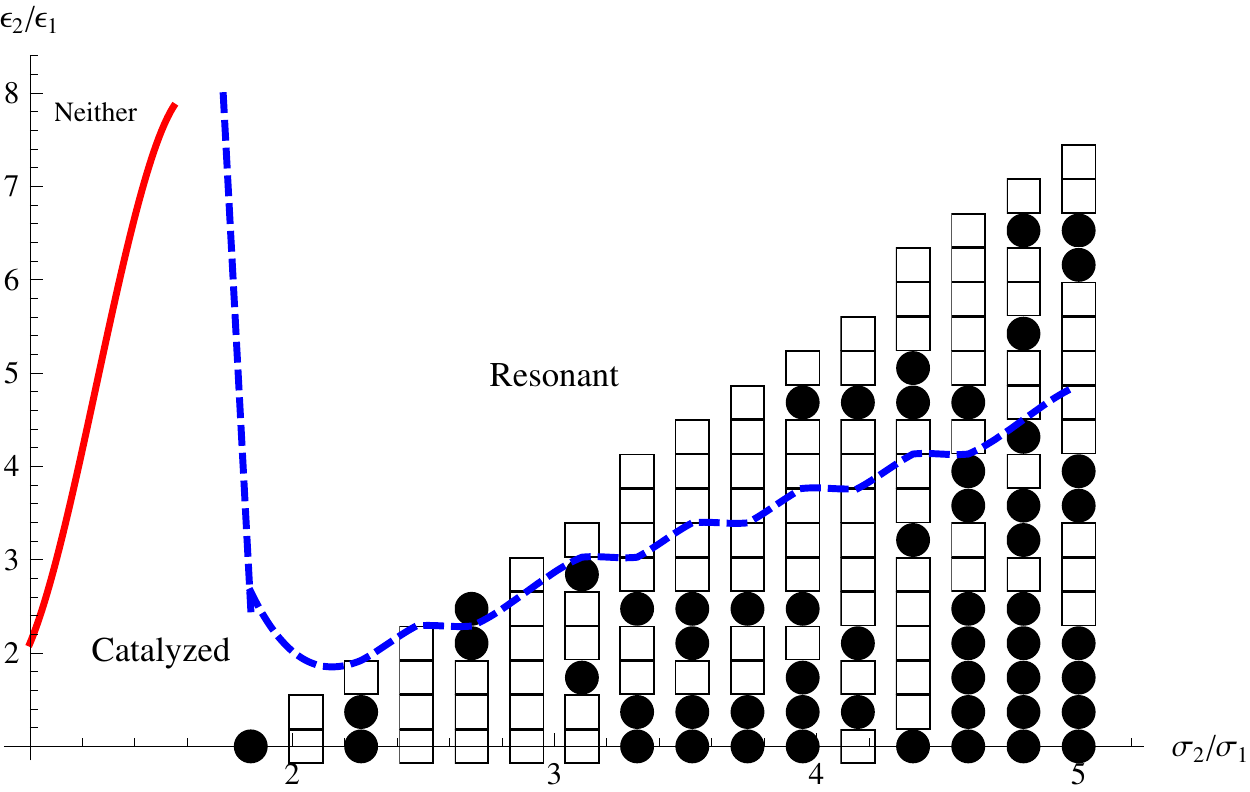}
\caption{This plot illustrates for which regions in parameter space the resonant effect is important.  Here $\epsilon_1 = 1.0 \cdot 10^{-4} {\rm J / m^3}$ and $\sigma_1 = 2.5 \cdot 10^{-11} {\rm J / m^2}$ are fixed (with $v_F = 55 {\rm m/s}$ so $S_E^{A \to A'} / \hbar \sim 8 \cdot 10^3$) and a linearly spaced grid of points are sampled.  To the left of
the red solid line, neither catalyzed nor resonant tunneling can occur.  The blue dashed line divides the region in which
catalyzed tunneling occurs from the region in which resonant tunneling occurs.
Both lines are approximate and are based on numerical simulations.  The centers of the squares
indicate points at which $100 < -S_E^{A \to A'} / (\hbar \ln P) < 1000$ and the centers of the
circles indicate points at which the tunneling is essentially unsuppressed: $1000 < -S_E^{A \to A'} / (\hbar \ln P)$, where $P$ is $P(A \to B)$ in the resonant tunneling regions and $P$ is $P(A \to A^\prime)$ in the catalyzed tunneling regions.
In general the enhancement due to the resonance effect becomes more likely as $\epsilon_2/\epsilon_1$ decreases and as $\sigma_2/\sigma_1$ increases.  The peaks in $P$ are generally very narrow.}
\label{rate_plot}
\end{center}
\end{figure}

The smallest allowed $r_{2,\rm{min}}$ occurs where $\Phi = 0$ (\ref{phi}), and $\Phi$ increases monotonically with $r_2$ in the allowed region $r_{2,\rm{min}} < r_2 < r_{2,\rm{max}}$.  If
 $\Phi(r_{2,\rm{max}}) > \Theta(r_{2,\rm{max}})$ (\ref{Theta1}), there must be some $r_2^*$ for which $\Phi(r_2^*) = \Theta(r_2^*)$ because $\Theta$ is always positive.  If $W(r_2^*)=
 (n+1/2) \pi$ (\ref{Wdef}), then the tunneling probability will be approximately unity.  Generically the dominant contribution to the tunneling probability will come from either the closest
 point above $r_2^*$ or the closest point below $r_2^*$ on the energy conservation curve that satisfies $W = (n+1/2) \pi$.  Alternatively if $\Phi(r_{2,\rm{max}}) \ll
 \Theta(r_{2,\rm{max}})$, then there is no possibility of a large enhancement.

In the allowed region $W(r_2)$ decreases monotonically.  Holding $\epsilon_1/\epsilon_2$ and $\sigma_1/\sigma_2$ fixed, the range of $W$ increases as the single barrier instanton action $S_E^{\rm{A} \to \rm{A^\prime}} = \frac{27 \pi^2}{2} \sigma_1^4/(\epsilon_1^3 v_F)$ increases.  The shape of the curves in Figure \ref{nconstraint} do not change as $S_E^{\rm{A} \to \rm{A^\prime}}$ is increased, but as the range of $W$ increases it becomes more likely for a point satisfying $W = (n+1/2) \pi$ to occur very close to $r_2^*$.  Thus the probability of a large enhancement due to resonance effects tends to increase as single-barrier tunneling becomes more unlikely.

Let us now determine for which potentials $V(\phi)$ the resonant effect is important:
\begin{itemize}
\item We can use the quantity $-S_E^{A \to A'} / (\hbar \ln P)$ where $P$ is $P(A \to B)$ if resonant tunneling occurs and $P$ is $P(A \to A^\prime)$ if catalyzed tunneling occurs to estimate the presence of the resonant effect.  In the total absence of the resonant effect,
\e
 -S_E^{A \to A'} /(\hbar \ln P) \simeq \frac{S_E^{A \to A'}}{S_E^{A \to A'} + S_E^{A' \to B}} <1
 \q
 while $ -S_E^{A \to A'} / (\hbar \ln P) > 1000$ when the resonant effect begins to eliminate the exponential suppression factor in the tunneling rate. This is shown in Fig. \ref{rate_plot}. The center of each black dot or circle satisfies $-S_E^{A \to A'} /(\hbar \ln P) > 1000$, i.e., each black dot contains a region with $-S_E^{A \to A'} / (\hbar \ln P) > 1000$.
 \item Within each dot, there may be points where the resonant effect is significantly more pronounced and $P \sim 1$.
 \item If we enlarge the plot (Fig. \ref{rate_plot}) to a three-dimensional plot, with $S_E^{A \to A'}$ as the third axis, we expect that there are points within each three-dimensional cluster of dots and each isolated dot where $P \sim 1$. If there are a number of $A'$ sub-phases available, nature will automatically pick the one with the fastest tunneling rate for the $A \to B$ transition.

\item As shown in Fig. \ref{rate_plot}, large enhancements in the tunneling probability due to resonant effects can only occur in certain regions of the $(\sigma_1, \sigma_2, \epsilon_1, \epsilon_2)$ parameter space.  Of particular interest is the $\epsilon_2 \gg \epsilon_1$, $\sigma_2 \gg \sigma_1$ limit, as this region supports resonant tunneling, has the possibility of a large enhancement in the tunneling probability, and describes superfluid Helium-3 near the transition temperature. If $\epsilon_2 \gg \epsilon_1$, $\sigma_2 \gg \sigma_1$, and $r_2 \sim r_1$, then $r_{2,\rm{max}}$ approaches $\lambda_{2c}$.  In the limit that the classically allowed region is shallow we see that
 $\Phi(r_{2,\rm{max}}) \gtrsim \Theta(r_{2,\rm{max}})$ only if
\e
\frac{(\sigma_2)^8}{\epsilon_2^6} \gtrsim (\sigma_1)^2 \bigg[\frac{(\sigma_1)^2}{\epsilon_1^2}-\frac{(\sigma_2)^2}{\epsilon_2^2}\bigg].
\q

\item However, as is clear from Fig \ref{rate_plot}, the resonant tunneling phenomenon persists in the region where both $\sigma_{2}/\sigma_1$ and $\epsilon_{2}/\epsilon_1$ increase. Keeping the $A'-B$ tension $\sigma_{2}$ and $A'-B$ energy density difference $\epsilon_{2}$ fixed, the resonant effect will be present (keeping the ratio $\sigma_{1}^4/\epsilon_{1}^3$ large enough to stay above the blue dashed line in Fig. \ref{rate_plot}) as both the
$A-A'$ tension $\sigma_{1}$ and the $A-A'$ energy density difference $\epsilon_{1}$ approach zero. Although the thin-wall approximation breaks down before we reach the limit, this does suggest that the resonant effect will remain in this limit under some appropriate conditions. Physically, it will mean that the coherence now comes from the  sum of paths in the degenerate or almost degenerate $A-A'$ phase.

\item In a physical system, as the barrier between $A$ and $A^\prime$ disappears, the tunneling probability from $A$ to $B$ must approach the tunneling probability from $A^\prime$ to $B$. That is, the system will simply roll from $A$ to $A'$ and then tunnel to $B$. If the barrier is too
small for a system at finite temperature, we expect that thermal effects will smear the resonant phenomenon.
\end{itemize}

\section{Some Predictions}

Let us first briefly review existing proposals to the fast $A \to B$ transition puzzle and then propose how our explanation may be tested. Although the predictions are very qualitative in nature, they are very distinctive. Some of the experiments proposed should be readily performed.

\subsection{Comparison to Other Explanations}

One well-known explanation of the rapid $A \to B$ transition is the ``Baked Alaska"  model \cite{baked, LY}. It proposes that the fast transition is triggered by cosmic rays, which provide the seeds of $B$ phase bubble nucleation. We do agree that showering ionizing radiation or shooting beams of external particles to a sample of $A$ phase superfluid He-3 can trigger the formation of $B$ phase nucleation bubbles, thus leading to a fast $A \to B$ transition \cite{Schiffer:1995zza}.  In the ``Baked Alaska" model,  an external disturbance such as a cosmic ray creates a localized hot region.  Instead of simply shrinking and disappearing, the localized hot region can evolve into a hot shell surrounding a cool interior \cite{Valls}.  In this model, the cold interior region cools sufficiently rapidly that it has a reasonable chance of being in the $B$ state, and the hot shell protects it from the surrounding $A$ phase until it grows larger than the critical size for expansion. 

Another well-known explanation of the rapid $A \to B$ transition is the ``cosmological" scenario \cite{BunkovTim, Bunkov}.  In this scenario after local heating, many casually independent regions independently undergo the superfluid transition to either the $A$ or $B$ phase.  If the $B$ phase seeds percolate, the transition can complete.  Because the regions independently undergo the transition, topological defects (in this case vortices) will be produced \cite{Kibble, Zurek}.  Measurements \cite{Bauerle} showing that not all of the energy is carried away by quasiparticles after neutron irradiation support this theory, since the missing energy is consistent with the energy expected to be stored in vortices.

\begin{figure}
\begin{center}
\mbox{\subfigure{\includegraphics[width=1.5in]{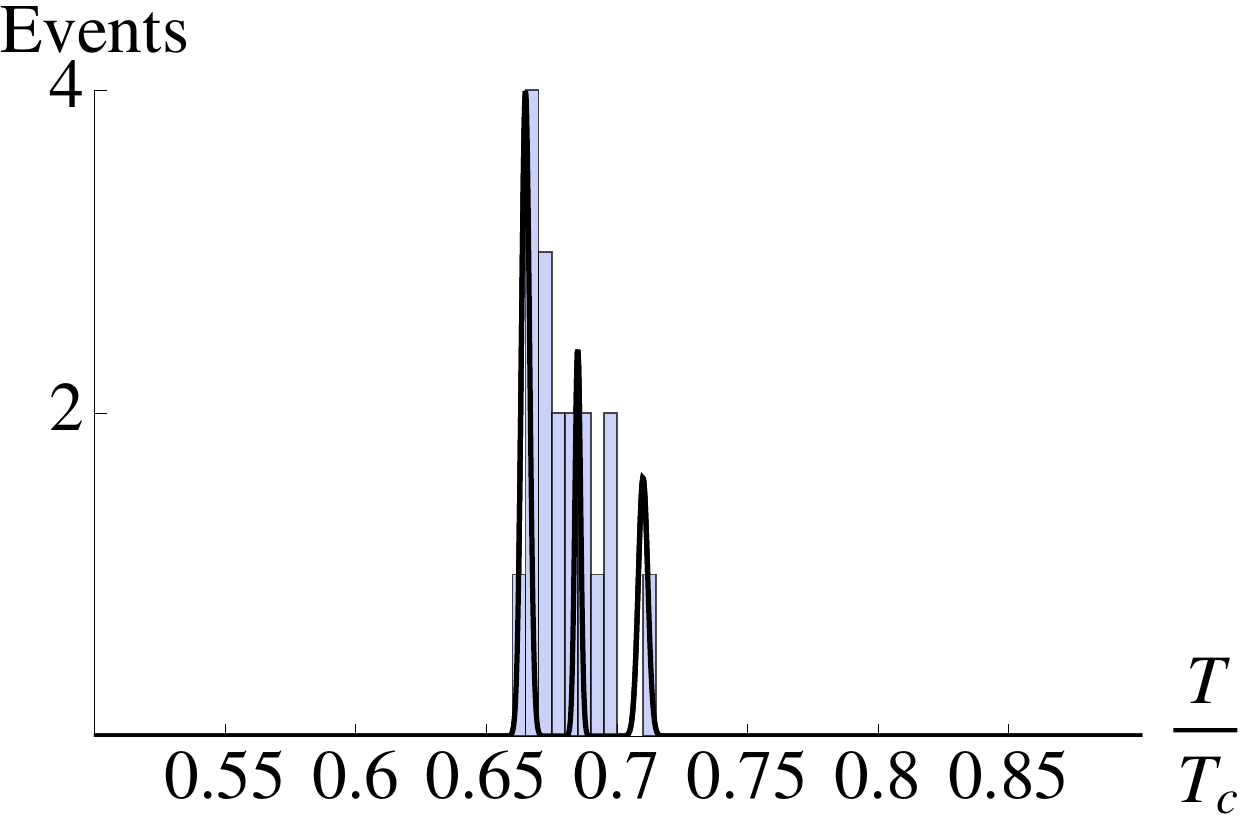}
\subfigure{\includegraphics[width=1.5in]{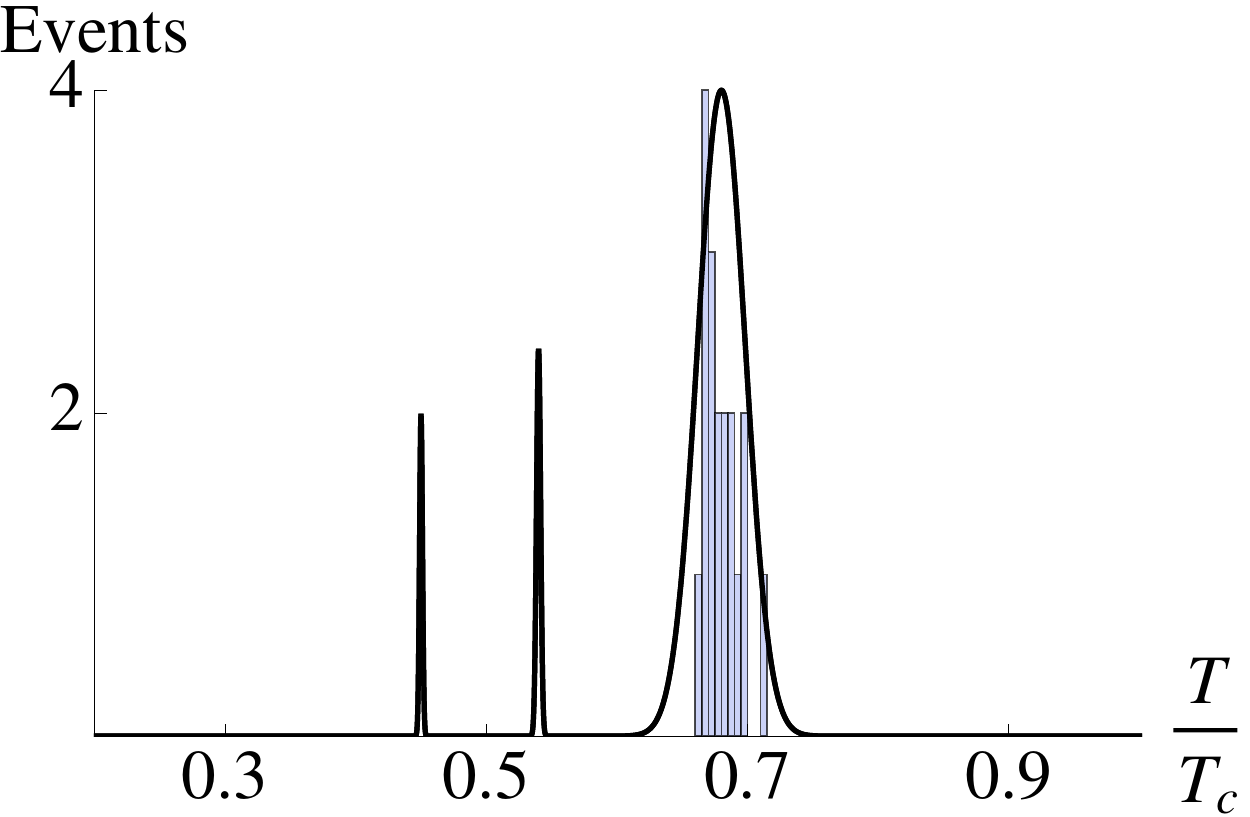} }}}
\caption{The number of $A \to B$ transition events in a $^3$He-$A$ sample as the temperature is slowly decreased. In the left figure, we illustrate the possibility that the broad resonant peak in Ref.\cite{Hakonen} is actually a collection of three unresolved narrow peaks at three nearby nucleation temperatures around $T=0.67T_c$.
The right figure assumes the broad peak at $T=0.67T_c$ is a single resonant peak (its breadth due to finite temperature effect) and there are additional peaks at nucleation temperatures below $T=0.67T_c$. These two features are not mutually exclusive.  Note that the expected number of events at a given temperature in this experiment is not proportional to $P(A \to B)$ at that temperature, because only a fraction of the trials reach the lower temperatures.  In both figures the bars indicate the number of observed $A$ to $B$ transitions in Ref.\cite{Hakonen}.}
\label{predictplot}
\end{center}
\end{figure}

Despite the successes of these theories, experiments have not ruled out transitions in the absence of cosmic rays.  As pointed out in Ref.\cite{Swift}, no correlation has been detected between nucleation events and coincidence counts from the cosmic ray detectors. Even though the cosmic ray detectors apparently do not cover all angles, this experiment strongly indicates that the fast $A \to B$ transition happens even in the absence of cosmic rays. That is the puzzle.

Another experiment \cite{LANL} shows that the nucleation temperature (about $0.67T_c$) depends on pressure and the geometry of the sample, again suggesting that cosmic rays are not the cause of the fast
 $A \to B$ transition.

These two experiments are in accord with our explanation, which has nothing to do with external disturbances.
One can certainly extend the cosmic ray detection to all angles to improve the experiment of Ref.\cite{Swift} to rule out with certainty that cosmic rays are not the reason for the fast $A \to B$ transition.

Vortex nucleation experiments in superfluid He-3 do not provide any direct test of our proposal because vortices are defects and hence the analysis of Section \ref{Review}, which relied on the formation of bubbles, is not readily applicable to the nucleation of vortices.  Vortex nucleation can occur in one of three ways: when the barrier disappears \cite{Parts}, when the sample is stimulated by external radiation \cite{Ruutu}, or when shear instability occurs at the $A$-$B$ interface \cite{Blaauwgeers}.  
The second of these explanations is the ``cosmological" scenario discussed above.  If the first or third of these explanations can be applied to $B$ phase nucleation, the experimental signatures will be distinct from those we describe in the following section.

\subsection{A Plausible Prediction}

The nucleation temperature is simply the temperature that the $A \to B$ transition takes place. In our scenario, that means the nucleation temperature satisfies the resonant condition. Slightly away from the resonant condition, the transition time becomes exponentially long.
Since the resonant tunneling phenomenon requires the satisfaction of a fine tuned condition, $A \to B$ transition happens only at specific values of pressure, temperature and magnetic field. Move slightly away from those values and the transition simply will not happen.
Viewed another way, a small change in geometry, external magnetic field or pressure will certainly shift the properties of the various He-3 phases, so the nucleation temperature  will be shifted accordingly. This seems to be the case in Ref.\cite{LANL,Hakonen}.

Consider the nucleation temperature already seen in Ref.\cite{LANL,Hakonen}. For example, Ref.\cite{Hakonen} finds that, for a $^3$He-$A$ sample with pressure at $29.3$ bar and magnetic field $H=28.4$ mT, $B$ phase nucleation takes place at temperature $T =0.67T_c$ with a (full) width of about $0.02T_c$ when the sample is slowly cooled, at a rate of $5  \mu {\rm K}$ per minute to $29  \mu {\rm K}$ per minute.  At this magnetic field and pressure, the $A$ and $B$ phases have equal free energy density at a temperature $T_{AB}=0.85T_c$.  In the event number (of $A \to B$ transitions) versus temperature plot (collected over a number of these temperature sweeps), the event number shows up as a resonance peak at  $T =0.67T_c$.  The dependence of the transition probability $P(A \to B)$ on temperature can be inferred from the number of observed transitions at each temperature and the cooling rate, although not from either individually.  Such a resonance peak is in accordance with our expectations. One can perform a more detailed data collection in the three-dimensional plot of the pressure, magnetic field and temperature to find the regions where the  $A \to B$ transition happens. We expect multiple isolated regions. These regions may take the form of isolated peaks, or lines or surfaces in the 3-dimensional space with temperature, pressure, and magnetic field as the three coordinates.

Since the resonant condition is simply the Bohr-Sommerfeld quantization condition (\ref{resonancecond}) and there are multiple solutions to this condition, there should be more than a single nucleation temperature.
That is, for fixed pressure and magnetic field, there may be additional nucleation temperatures. In Fig. \ref{predictplot}, we present two plausible scenarios of multiple nucleation temperatures :
\begin{itemize}
\item The peak in the event number (of $B$ phase transition) versus temperature in Ref.\cite{Hakonen} is actually an unresolved collection of two or more very narrow peaks. This is illustrated in the left panel of Fig. \ref{predictplot}. To resolve them and to determine their actual widths, one may have to use a slower rate in the temperature sweep and collect more data.
If the finite temperature effects are small, then the width of each individual peak can easily be much less than $1 \mu {\rm K}$. Since the width of the unresolved peak is about $50 \mu {\rm K}$, sitting at a random fixed temperature within this broad width is unlikely
to encounter an $A \to B$ transition (once off the peak, the transition probability becomes exponentially small). This agrees with the observation of Ref.\cite{Hakonen}, that a $^3$He-$A$ sample can sit at a stable temperature in that temperature range for hours.

\item The width of the peak may be due to the spread caused by the finite temperature and other effects.
In this case, there can be other critical nucleation temperatures besides the one observed. This is illustrated in the right panel of Fig. \ref{predictplot}. A simple search of additional nucleation temperatures below $T =0.67T_c$ will be very interesting. If they exist, we expect their widths to be narrower as well.
\end{itemize}

Although we do not know enough about the detailed structure of the free energy functional to find the positions or shapes of the additional resonances, we can still make some comments here :
\begin{itemize}
\item Our prediction only states that we expect more than a single nucleation temperature or a single resonance peak. That is, we predict the existence of narrow peaks in the transition rate, between which the transition is completely absent.  The positions and shapes of the narrow resonance peaks shown in Fig. \ref{predictplot} are not predictions, and are shown for illustrative purposes only.
\item It is entirely possible that the actual scenario incorporates both features of  narrow peaks just described.
\item Ref.\cite{Hakonen} sees the nucleation temperature only during cooling down, not during warmup. This suggests that the warmup nucleation temperature may be shifted outside the temperature range studied. Another possibility may be the tunneling probability is still too low even when hitting the resonance condition.
As pointed out in Ref.\cite{Hakonen}, this difference between cooling down and warmup may be due to the continuous vortices induced by rotation in the sample of $^3$He-$A$.
\item It is possible that the additional resonances show up more readily if one adjust slightly the pressure
and/or the magnetic field.
\item Although we generically expect to have multiple resonances with different quantization number in the quantization condition (\ref{resonancecond}), it is possible that variation of temperature, pressure, or magnetic field leads to variations of the effective potential $V(\phi)$ (in Fig. \ref{double_barrierV}) for the interpolating field $\phi$. This in turn leads to corresponding variations in $V(\lambda)$ (in Fig. \ref{VmU}) and so $\Theta$, $\Phi$ and $W$ in Appendix A in a way such that two or more distinct resonances appear at the same quantization number. We cannot rule out the possibility of such a coincidence.
\end{itemize}

\subsection{Growth of Bubbles}

Suppose for a particular choice of parameters, resonant tunneling occurs.  After nucleation, both bubbles will grow classically.  By symmetry, the radii $|\bf{x}_1|$ and $|\bf{x}_2|$ of the two bubbles satisfy
\begin{equation}
r_i^2 = |\bf{x}_i|^2 - t^2
\end{equation}
where $i=1,2$ assuming both bubble walls have zero velocity at $t=0$.  Thus
\begin{equation}
\frac{|\bf{x}_2|^2}{|\bf{x}_1|^2} = \frac{r_2^2+t^2}{r_1^2+t^2}
\end{equation}
which implies that smaller bubble always grows faster, since this ratio is monotonic in $t>0$, and approaches unity at future infinity.  In a physical system, the bubble walls will interact, and once they are close enough they will merge.  In superfluid He-3, energy dissipation could complicate this simple treatment, but should not alter the conclusion that the separation between the two bubble walls is initially decreasing.

The distance between the two walls at nucleation could be much larger than the thickness of either individual wall.  If it were possible to observe the nucleation of the bubbles without sufficient resolution to separate the double walls, they will appear as a single thick wall. Then we expect to see that ``thick wall" becomes thinner as the bubble grows.

\section{Discussion and Remarks}

Now, we like to summarize the scenario we envision. In a typical experiment trying to reach the $B$ phase of superfluid He-3, the sample starts at the $A$ phase, say, the $A^i$ sub-phase. One reaches this phase via either the normal or the $A_1$ phase, where the $A_1 \to A$ transition is second order. As the temperature is lowered to supercool the $A^i$ sub-phase, fast tunneling requires the presence of a $A^j$ sub-phase slightly below the $A^i$ phase.

As the temperature of the sample is being lowered (in some experiments, adjustment of pressure and/or external magnetic field may also take place), all the properties (say $\sigma_{1,2}$ and $\epsilon_{1,2}$) will be varying slowly. This fine sweeping of the parameters of the system (as well as the choices of $A^j$ and $B^k$) offers a good chance that resonant tunneling with vanishing (or almost vanishing) exponent will be hit at certain point for specific choices of the $A^j$ and $B^k$ sub-phases., enabling tunneling with little or no exponential suppression. This scenario clearly requires the existence of the $A^j$ sub-phase.

Note that we are not concerned with tunneling from the $B$ phase back to the $A$ phase. Presumably, even in the $B$ phase, some regions adjacent to the container walls will remain in the $A$ phase, so that when the temperature is raised so that the $A$ phase becomes the true ground state, those $A$ phase regions will simply grow and take over the sample.

\subsection{Some Subtleties}

Notice that our analysis assumes homogeneity and isotropy of the medium. However, many sub-phases are not homogeneous and/or isotropic. Explicit calculations of the tunneling in such situations will be much more complicated. However, one may convince oneself that resonant tunneling is a generic phenomenon, independent of the details, as long as some constraints are satisfied; that is, the presence of a classically allowed region that allows the Bohr-Sommerfeld quantization condition (i.e., the coherent sum of Feynmann paths) to be satisfied. As we have seen, this is not a very tight constraint when the $A^j$ sub-phase is present.
Experimentally and/or theoretically, one has to check that such a sub-phase is actually present. This is a qualitative prediction.

Additionally our analysis neglects thermal fluctuations.  On general grounds we expect thermal effects to broaden the resonances, but if $S_E^{A \to A'} / \hbar$ is sufficiently large these effects are likely negligible.

\subsection{Cosmic Landscape}

In superstring theory, we believe there are classically stable local vacua, described by many ``parameters and variables" known as moduli. They number in the dozens to hundreds. In superfluid He-3, we have a complex $3 \times 3$ matrix as the order parameter plus many interaction parameters. (Here, tiny interaction terms can be important in reaching the sweet spots of resonant tunneling.) Not surprisingly, both systems have many solutions : classically stable local vacua in string theory, collectively known as the cosmic landscape, or phases in He-3. That a phase transition in He-3 is much much faster than naively expected is a pleasant surprise for experimentalists. This phenomenon should be fully understood so we can decide whether the same phenomenon should happen in the cosmic landscape. Here we speculate that, due to the resonant tunneling effect, the tunneling transitions in the cosmic landscape may happen surprisingly fast. In fact, the transitions may simply become exponentially faster as the number of vacua becomes more numerous.  Resonant and catalyzed tunneling could be relevant to eternal inflation (see \cite{Guth:2007ng} for a review). The implication of this phenomenon on the behavior of the universe cannot be understated.  It is interesting that He-3 experiments may help clarify some outstanding theoretical issues in cosmology.

\acknowledgments

We thank Jeevak Parpia who drew our attention to the fast tunneling puzzle in superfluid He-3.  We thank Chris Henley for detailed comments on a draft. We thank Jason Ho, Doug Osheroff and Grisha Volovik and our He-3 colleagues at Cornell for valuable discussions.
This work is supported in part by the National Science Foundation under grant PHY-0355005.

\appendix

\section{Resonant Tunneling in Quantum Mechanics}

Now let us briefly review resonant tunneling in quantum mechanics.  We consider a particle with a unit mass moving under the influence of a one-dimensional potential $V(\lambda)$. Using the WKB approximation to solve the one-dimensional time-independent Schr\"odinger equation (\ref{QMl})
for the wavefunction of the particle $\Psi_0(\lambda)$ gives the linearly
independent solutions
\e
\psi_{L,R}(\lambda) \approx \frac{1}{\sqrt{k(\lambda)}} \exp \bigg( \pm i \int d\lambda k(\lambda) \bigg)
\q
in the classically allowed region, where $k(\lambda) = \sqrt{\frac{2m}{\hbar^2}(-V(\lambda))}$, and
\e
\psi_{\pm}(\lambda) \approx \frac{1}{\sqrt{\kappa(\lambda)}} \exp \bigg( \pm \int d\lambda \kappa(\lambda) \bigg)
\q
in the classically forbidden region, where $\kappa(\lambda) = \sqrt{\frac{2m}{\hbar^2}(V(\lambda))}$.  A complete solution is given by $\psi(\lambda) = \alpha_L \psi_L(\lambda)+\alpha_R \psi_R(\lambda)$ in the classically
allowed region and $\psi(\lambda) = \alpha_+ \psi_+(\lambda)+\alpha_- \psi_-(\lambda)$ in the classically forbidden region.


We consider $V(\lambda)$ with three classically allowed regions as shown in Fig. \ref{VmU}, where $V(\lambda)=0$ for $\lambda <0$. The same analysis gives the tunneling probability from ${\tilde A}$ to ${\tilde B}$, via ${\tilde A'}$, as \cite{Merzbacher,Tye:2006tg},
\begin{eqnarray}\label{A_to_C}
P({\tilde A} \to {\tilde B}) = 
\nonumber \\ 4 \left( \left( \Theta \Phi + \frac{1}{\Theta \Phi}\right)^2 \cos ^2 W + \left( \frac{\Theta}{\Phi} + \frac{\Phi}{\Theta} \right)^2  \sin^2 W \right)^{-1}\,\,,
\end{eqnarray}
where
\e
\label{Theta1}
\Theta \simeq 2 \exp \left(\frac{1}{\hbar} \int_{0}^{\lambda_A} d\lambda \sqrt{2V(\lambda)} \right) \,\, ,
\q
and
\e
\label{phi}
\Phi \simeq 2 \exp \left( \frac{1}{\hbar} \int_ {\lambda_{\Lambda}}^{r_1} d\lambda \sqrt{2V(\lambda)}  \right)
\q
are typically exponentially large, and
\e
\label{Wdef}
W = \frac{1}{\hbar}\int_{\lambda_A}^{\lambda_{\Lambda}} d\lambda  \sqrt{ -2V(\lambda)}\,\, ,
\q
with $\lambda_{\Lambda}$ and $r_1$ the turning points on the barrier between ${\tilde A'}$ and ${\tilde B}$.
Here, the region ${\tilde A'}$ ($\lambda_A <\lambda \le \lambda_{\Lambda}$) is classically allowed.

If ${\tilde A'}$ has zero width (i.e., this classically allowed region is absent), $W=0$ so $P({\tilde A} \to {\tilde B})$
is very small,
\bea
P({\tilde A} \to {\tilde B}) \simeq 4 \Theta^{-2} \Phi ^{-2} =
P({\tilde A} \to {\tilde A'})P({\tilde A'} \to {\tilde B})/4
\eea
However, if $W$ satisfies the Bohr-Sommerfeld quantization condition for the $n$th bound state in ${\tilde A'}$, namely
\bea
W=(n+1/2) \pi, \quad \quad n=0,1, 2, . . .
\label{resonancecond}
\eea
then $\cos W =0$, and the tunneling probability approaches a small but not necessarily
exponentially small value
\bea
P({\tilde A} \to {\tilde B})  = \frac{4}{\left(\Theta/\Phi+ \Phi/\Theta \right)^{2}}
\label{resonancetnn}
\eea
This is the resonance effect. If $\Theta \to \Phi$,
$P({\tilde A} \to {\tilde B}) \rightarrow  1$, that is, the tunneling probability approaches unity.  Notice that the existence of the resonant tunneling effect
here is independent of the details, though some fine-tuning may be necessary. In superfluid He-3, we argue that the choices of the specific $A$ and $B$ sub-phases plus the slowly changing environment (say, changing temperature) provides some natural effects that mimic the fine-tuning needed.

The above phenomenon is easy to understand in the Feynman path integral formalism. A typical tunneling path starts at ${\tilde A}$ and tunnels to ${\tilde A'}$. It bounces back and forth $k$ times,
where $k=0,1,2,... \infty$, before tunneling to ${\tilde B}$. When the Bohr-Sommerfeld quantization condition (\ref{resonancecond}) is satisfied, all these paths interfere coherently,
leading to the resulting resonant tunneling phenomenon.
It is important to point out that this constructive interference effect cannot be captured in a pure Euclidean formulation typically used in a quantum tunneling problem.


\end{document}